\documentclass[floatfix,twocolumn,preprintnumbers,amsmath,amssymb,prb]{revtex4}
\usepackage{graphicx}
\usepackage{dcolumn}
\usepackage{bm}
\usepackage{psfrag}

\graphicspath{{./}}

\newcommand{\be}{\begin{equation}}
\newcommand{\ee}{\end{equation}}

\newcommand{\br}{{\mathbf{r}}}

\begin{document}

\title{On the spatial structure of anomalously localised states in disordered conductors}
 \author{V. Uski}
\author{B. Mehlig} 
\affiliation{Physics and Engineering Physics, Gothenburg University/Chalmers, SE-41296 Gothenburg, Sweden
}
\author{M. Schreiber}
\altaffiliation[On leave from: ]{Institut f\"ur Physik, Technische Universit\"at, D-09107 Chemnitz, Germany}
\affiliation{School of Engineering and Science, International University Bremen, D-28725 Bremen, Germany}

\date{\today} 
\begin{abstract} 
We study the spatial structure of wave functions with exceptionally high
local amplitudes in the Anderson model of localisation. By means of exact
diagonalisations of finite systems, we obtain
and analyse images of these wave functions:
we 
compare the spatial structure of such anomalously localised states
in quasi-one-dimensional samples to that in
three-dimensional samples. In both cases the average wave-function intensity
exhibits a very narrow peak. The background intensity, however, is found
to be very different in these two cases: in three dimensions, 
it is constant, independent of the distance to the localisation centre
(as expected for extended states). In quasi-one
dimensional samples, on the other hand, 
it is redistributed 
towards the localisation centre and approaches a characteristic
form predicted in~[A. D. Mirlin, Phys. Rep. {\bf 326}, 249 (2000)].
\end{abstract}

\maketitle 

Statistical properties 
of physical observables in disordered electronic
quantum systems have attracted considerable interest in the last decade.  In
such  systems, quantum interference may cause  the (non-interacting)
conduction electrons to localise\cite{And58}. In three dimensions (3D), this
Anderson localisation occurs when the disorder strength exceeds a critical
value.  Beyond this value (which depends on the Fermi energy and the symmetries
of the system), electron wave functions are confined to a limited spatial
region of the sample.  

In the metallic regime, on the contrary,  wave functions typically
spread over the whole sample, and they contribute to electron transport.
However, some wave functions show localised behaviour even in this weakly
disordered regime (these are so-called anomalously localised states, abbreviated as ALS
in the following).
This leads to a non-zero, albeit small,
probability of observing exceptionally large wave-function amplitudes,
often in the form of a log-normal distribution function of wave-function
intensities.
ALS in electronic conductors have been studied intensively in recent years,
using the so-called diffusive non-linear sigma model
(DNLSM)\cite{MirF93,FyoM94,FalE95b}.  An overview of the main results and
predictions based on the DNLSM is given in Ref.~\onlinecite{Mir00}. 
Moreover, possible complications due to non-diffusive,
so-called ballistic effects on length scales smaller than the mean free path
are  discussed.  As was pointed out in Ref.~\onlinecite{SmoA97}, these may modify the predictions
of the DNLSM (see also Refs.~\onlinecite{BlaMM01} and \onlinecite{UskMS01}).
ALS are expected to occur in lower-dimensional disordered systems, too, when the disorder is
weak.  

These interesting 
analytical results have motivated a number of numerical studies:
in Ref.~\onlinecite{MulMMS97}, for example,
log-normal statistics of wave-function
amplitudes in two-dimensional (2D) conductors near the delocalisation-localisation 
transition was observed. In Ref.~\onlinecite{UskMRS00} it was confirmed that 
as the disorder is reduced to reach the weakly disordered regime,
the distribution function remains log-normal. It has, however, not been
possible to resolve a 
discrepancy (between the DNLSM\cite{Efe83,Mir00} and the so-called
direct optimal
fluctuation method\cite{SmoA97}) in the prediction of
the parameters of this distribution.
As far as the numerical work is concerned,
the situation may be summarised as follows.
While it can be concluded that the DNLSM appears to be adequate in
the quasi-one-dimensional (Q1D) Anderson model under certain
conditions\cite{UskMRS00,UskMS01}, the DNLSM may not correctly
describe ALS in two and three dimensions\cite{UskMS01,Nik02}, at least for the
parameter values considered in these studies. On the other hand, it was found
recently \cite{OssKG02} that the DNLSM appears to describe the statistics of
rare events adequately in a 2D kicked rotor.  

A reason for the possible failure of the DNLSM to describe ALS in
the 3D Anderson model may be the importance of short length
scales\cite{SmoA97,BlaMM01}, see also Ref.~\onlinecite{UskMS01}.
The DNLSM is based on the semiclassical
picture of a diffusing electron in a random potential. In this
picture the smallest relevant length scale is the electronic mean
free path $\ell$ between elastic collisions.  Therefore the DNLSM
cannot describe situations where length scales smaller than $\ell$
are important. Such a situation could occur if ALS were created
not through semiclassical diffusion, but through local potential
wells trapping the electrons\cite{SmoA97,new}.
The mechanism giving rise to ALS is expected\cite{SmoA97,Mir97,Mir00} to
crucially depend on the dimensionality of the system and may determine their
spatial
structure.  

Finally, within the DNLSM, it is possible to obtain the wave-function
statistics directly in Q1D, using a transfer-matrix
technique. In 2D and 3D, on the other hand,
a further saddle-point approximation is necessary.  

Given this situation, further numerical work describing
ALS in disordered conductors is called for. In the following,
we describe results of exact diagonalisations of finite
Anderson models of localisation, yielding averages of the 
{\em spatial structure} of ALS. While this spatial
structure has been studied in detail within the DNLSM~\cite{Mir97,Mir00},
numerical imaging of wave functions with anomalously amplitudes
has not yet been performed~\cite{Nik01}.

The Anderson model is defined by the tight-binding Hamiltonian
\begin{equation}
\label{eq:defH}
\widehat{H} =  \sum_{\br,\br'} t^{\phantom \dagger}_{\br\br'} c_\br^\dagger \
c^{\phantom\dagger}_{\br'}
+ \sum_\br {\upsilon}^{\phantom\dagger}_\br c_\br^\dagger 
c^{\phantom\dagger}_\br
\end{equation}
on a hypercubic lattice.
Here $c_\br^\dagger$ and $c_\br$ are the creation and annihilation
operators at site $\br$, 
the hopping amplitudes are $|t_{\br\br'}|=1$ for nearest-neighbour 
sites and zero otherwise.  The on-site potentials 
$\upsilon_\br$ are Gaussian distributed with zero mean and
$\langle\upsilon_\br \upsilon_{\br'}\rangle = (W^2/12)
\delta_{\br\br'}$.
As usual, the parameter $W$ characterises the strength of the
disorder and $\langle\cdots\rangle$ denotes the disorder average. 
We study finite Q1D and 3D lattices.

It has been suggested in Ref.~\onlinecite{FyoM94} to characterise the
spatial structure of ALS by means of conditional averages of the form
$\left\langle V^q\vert\psi(\br)\vert^{2q}\right\rangle_t\,$.
Here $\langle\cdots\rangle_t$ denotes an average
over all wave functions with $t=V\vert\psi(\mathbf{0})\vert^2$, 
$q=1,2,\dots$,
and $V$ is
the volume of the system
(wave functions are normalised
so that $\langle |\psi_j(\br)|^2\rangle = V^{-1}$).
For large values of $t$,
the average $\left\langle V^q\vert\psi(\br)\vert^{2q}\right\rangle_t$
describes how wave functions decay, on average,
away from the localisation centre.

In the metallic          regime, typical wave functions fluctuate as
described by random matrix theory\cite{Meh90,Haa92} (RMT).  Depending
on
the symmetries of (\ref{eq:defH}), 
Dyson's Gaussian orthogonal or
unitary ensembles are appropriate\cite{Meh90}. We refer to these two
cases by assigning, as usual, the parameter $\beta=1$ to the former
and $\beta = 2$ to the latter.
Within RMT, for $\br\neq 0$,
\begin{equation}
\label{eq:rmt}
\left\langle V^q\vert\psi(\br)\vert^{2q}\right\rangle_t=
\begin{cases}
1 & \text{if } q=1\,,\\
3 & \text{if } q=2\text{ and }\beta=1\,,\\
2 & \text{if } q=2\text{ and }\beta=2\,.
\end{cases}
\end{equation}
This reflects that in a metallic system, wave
functions spread uniformly over the whole sample 
with spatially short-ranged correlations\cite{FyoM94,BlaMM01}.
In the presence of ALS, the conditional averages
$\left\langle V^q\vert\psi(\br)\vert^{2q}\right\rangle_t$
are expected to differ from (\ref{eq:rmt}).

In order to characterise the spatial structure of ALS in
the weakly disordered Anderson model, the
conditional averages $\langle\cdots\rangle_t$ were calculated within
the DNLSM in Refs.~\onlinecite{FyoM94} and \onlinecite{Mir97}. 
For $q=1$ the authors write
\begin{equation}
\label{eq:psi2full}
\langle V\vert\psi(\br)\vert^2\rangle_t=g(E,t;\br)/f(E,t),
\end{equation}
where $f(E,t)$ is defined as
\begin{equation}
\label{eq:defP}
f(E,t) 
=\Delta\Big\langle\sum_j\!
\delta(t\!-\!V|\psi_j(\mathbf{0})|^2)\Big  \rangle_{E_j\simeq E} \,.
\end{equation}
Here $\Delta$ is
the mean energy level spacing and $\langle \cdots\rangle$
denotes a combined disorder and energy average (over a small
interval of width $\eta$ centered around $E$).
The function $g(E,t;\br)$ is defined as
\begin{equation}
\label{eq:defQ}
g(E,t;\br) 
=\Delta\Big\langle\sum_j\!V\vert\psi_j({\mathbf{\br}})|^2
\delta(t\!-\!V|\psi_j(\mathbf{0})|^2)\Big \rangle_{E_j \simeq E}\,.
\end{equation}
Below, we take the $\delta$-functions in (\ref{eq:defP}) and (\ref{eq:defQ})
to be slightly broadened with small but finite $\gamma > 0$.

 In close vicinity of the localisation centre, ALS are found to exhibit
 a very narrow peak (of width less than $\ell$). The 
 expressions below apply for $r > \ell$ and thus describe
 the smooth background intensity, but not the sharp
 peak itself\cite{Mir97}. 
For $\beta=2$ one obtains for a Q1D conductor\cite{FyoM94,Mir97}
\begin{equation}
\label{eq:ft}
f(E,t)=  
\frac{{\rm d}^2}{{\rm d}t^2}
\left[{\cal W}^{(1)}(t/X,\tau_+){\cal W}^{(1)}(t/X,\tau_-)\right]\\
\end{equation}
and (assuming $\vert \br\vert\equiv r>\ell$)
\begin{equation}
\label{eq:q}
g(E,t;\br)\!=\!-\!X\!\frac{{\rm d}}{{\rm d}t}\!\!\left[\frac{{\cal W}^{(2)}\!(t/X,\tau_1,\tau_2)
{\cal W}^{(1)}\!(t/X,\tau_-)}{t}\right]
\end{equation}
where $\tau_+=(L-x)/\xi$, $\tau_-=x/\xi$, $\tau_1=r/\xi$ and
$\tau_2=(L-x-r)/\xi$. Here,
$X=L/\xi$, $L$ is the length of the sample, and $\xi$ is the
localisation length. Moreover, $x$ is the distance of the observation point
from the edge of the sample (c.f. Refs.~\onlinecite{FyoM94,Mir97}).
The function ${\cal W}^{(1)}(z,\tau)$ obeys the differential equation
\begin{equation}
\frac{\partial }{\partial \tau} {\cal W}^{(1)}(z,\tau)
=\Big(z^2 \frac{\partial^2}{\partial z^2}-z\Big) {\cal W}^{(1)}(z,\tau)
\end{equation}
with initial condition ${\cal W}^{(1)}(z,0) = 1$. The function
${\cal W}^{(2)}(z,\tau,\tau^\prime)$ obeys the same differential
equation, but with the initial condition 
${\cal W}^{(2)}(z,0,\tau) =z{\cal W}^{(1)}(z,\tau)$.

For large values of $t$,  it is suggested in Ref.~\onlinecite{Mir97}
that the background intensity should be given by
\begin{equation}
\label{eq:psi2asy}
\langle V\vert\psi(\br)\vert^2\rangle_t\approx\frac{1}{2}\sqrt{t
X}\left(1+r\sqrt{t/(L\xi)}\right)^{-2}
\end{equation}
where in accordance with the above $r > \ell $ is assumed,
and also $r \ll \xi$.
Corresponding expressions may be obtained
for higher values of $q$. One expects\cite{Mir97}
(for $\ell < r\ll\xi$)
\begin{equation}
\label{eq:psiq}
\langle V^q\vert\psi(\br)\vert^{2q}\rangle_t\simeq 
q!\left[\langle V\vert\psi(\br)\vert^{2}\rangle_t\right]^q.
\end{equation}

In 3D, by contrast, results corresponding to (8-10) are not known\cite{Mir97},
but it is expected that the background intensity of 
ALS is characteristically different from
that in Q1D samples: ALS in
3D systems are expected~\cite{SmoA97,Mir00} to exhibit a very 
narrow maximum near the localisation centre (the width of
this peak is a matter of current debate\cite{SmoA97,Mir00}).
The background, however, is expected~\cite{Mir97,Mir00} 
to decay very quickly  to
$\langle V \vert\psi(\br)\vert^{2}\rangle_t\simeq 1$.
Furthermore, fluctuations around this average are
expected\cite{Mir97} to be described by 
RMT [see Eq.~(\ref{eq:rmt})].

In the following we describe and discuss results of
exact-diagonalisations\cite{ElsMMR99} of the Anderson Hamiltonian
(\ref{eq:defH}) and compare to the results of the previous section. We
emphasise that at least in 3D it is important to use a Gaussian distribution
for the on-site potentials in order to be able to compare to the analytical
predictions: ALS are  possibly {\em non-universal}, their spatial 
structure may depend on the properties of the random potential. 

\begin{figure}[ht]
\centerline{\includegraphics[clip,width=6cm]{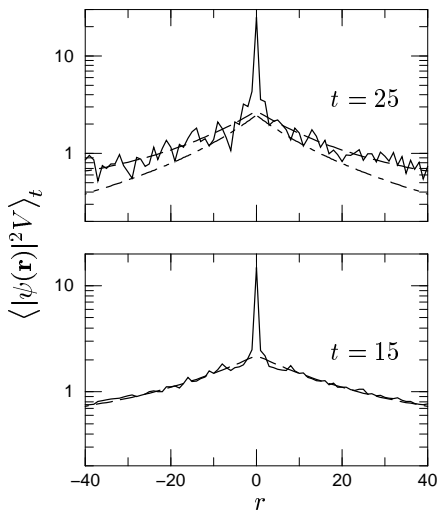}}
\caption{\label{fig:alsq1d} 
Spatial structure of anomalously localised wave functions with
$t=\vert\psi(0)\vert^2V$. Solid lines: Numerical results for $V=128\times
4\times 4$ (Q1D case), disorder $W=1.6$ and energy $E\simeq
-1.7$, averaged over 40 000 wave functions. Dashed lines: Analytical
predictions with $X=0.97$ [full formula, with Eqs.~(\ref{eq:ft}) and
(\ref{eq:q})]. The dash-dotted line 
shows the asymptotic formula, Eq.~(\ref{eq:psi2asy}), for $t=25$.}
\end{figure}

\begin{figure}
\centerline{\includegraphics[width=6.8cm]{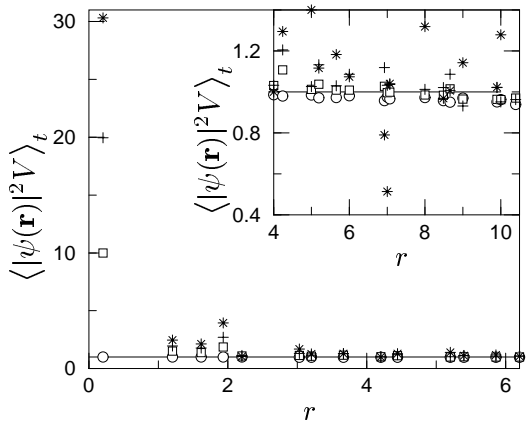}}
\caption{\label{fig:als3d} 
 Spatial structure of anomalously localised wave functions with
 $t=\vert\psi(0)\vert^2V$. Numerical results for 
 $V=48\times 48\times 48$ (3D
 case), disorder strength $W=2$ and energy $E\simeq -1.7$, averaged
 over 88 wave functions. The symbols show the numerically calculated
 values for $t=1$ ($\circ$), $t=10$ ($\Box$), $t=20$ ($+$),
 and $t=30$ ($*$).
The line shows the constant RMT average
 intensity $\langle\vert\psi(\br)\vert^2V\rangle_t=1$.
}
\end{figure}

\begin{figure}[ht]
\centerline{\includegraphics[clip,width=6cm]{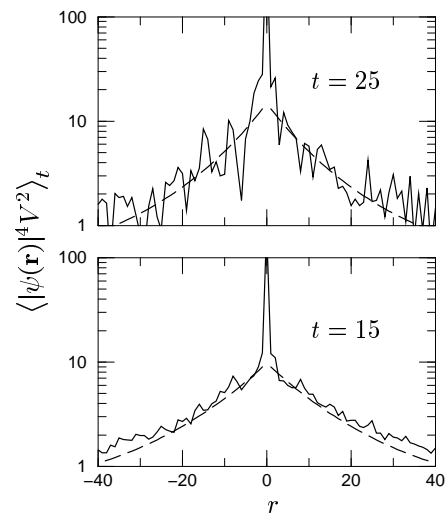}}
\caption{\label{fig:varq1d} 
Solid lines: The function $\left\langle V^2\vert
\psi(r)\vert^4\right\rangle_t$ in the Q1D case averaged over the
same wave functions as in Fig.~\protect\ref{fig:alsq1d}.  The dashed
lines show the function $2[\left\langle\vert
V\psi(r)\vert^2\right\rangle_t]^2$, where $\left\langle\vert
V\psi(r)\vert^2\right\rangle_t$ was calculated from eqs. (8-10).}
\end{figure}

\begin{figure}[ht]
\centerline{\includegraphics[clip,width=7.8cm]{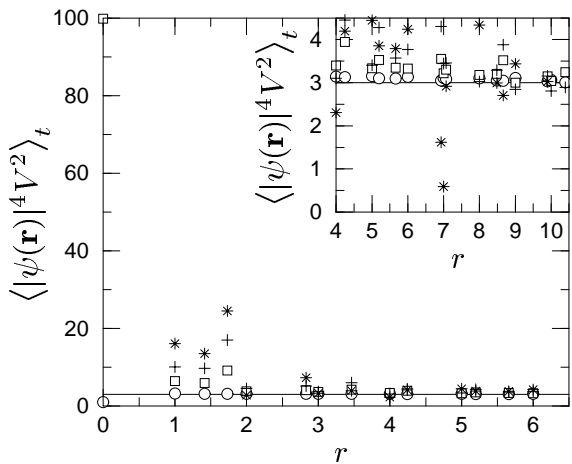}}
\caption{\label{fig:var3d} 
The function $\left\langle V^2\vert \psi(r)\vert^4\right\rangle_t$ in
the 3D case.  The symbols show the numerically calculated
values. The parameters and
symbols are the same as in Fig.~\protect\ref{fig:als3d}. The line shows the
constant RMT fluctuations, $\left\langle V^2\vert \psi(r)\vert^4\right\rangle_t=3$.}
\end{figure}

Our numerical results  for
$\langle V\vert\psi(\mathbf{r})\vert^2\rangle_t$ and $\langle
V^2\vert\psi(\mathbf{r})\vert^4\rangle_t$ 
are  summarised in Figures~\ref{fig:alsq1d}, \ref{fig:als3d},
\ref{fig:varq1d}, and \ref{fig:var3d} .
Figs.~\ref{fig:alsq1d} and \ref{fig:varq1d} show
results for Q1D samples with  $\beta = 2$,
while figs.~\ref{fig:als3d} and \ref{fig:var3d} show
results for 3D samples and $\beta = 1$.

In all cases we see a very narrow peak at the localisation
centre. Our main results, however, concern 
characteristic differences between
the distribution of background intensities in Q1D and 3D samples.
In the Q1D case we observe  a global redistribution of background
intensity towards the localisation centre
(as compared with uniformly spread RMT wave functions).
The numerical results are very well described
by Eqs.~(6-8). The asymptotic formula (9) considerably underestimates
$\langle V\vert\psi(\br)\vert^2\rangle_t$ for the values of $t$ used
in Fig.~\ref{fig:alsq1d}.
In the 3D case, by contrast, the background intensity is roughly
constant (Fig.~\ref{fig:als3d}). This 
is consistent with the qualitative picture summarised above.

Figs.~\ref{fig:varq1d} and \ref{fig:var3d} show the second moments, i.e., the
case $q=2$. In Q1D samples, Eq.~(\ref{eq:psiq}) appears to be  valid for large
$t$, as expected.  In 3D samples, the fluctuations of the background intensity
appear to be consistent with the RMT statistics (although the scatter is large).
That is, they are roughly constant as a function of $r$,  as predicted in
Ref.~\onlinecite{Mir97}. However, a closer inspection shows (inset of
Fig.~\ref{fig:var3d}) that the second moment is somewhat higher than expected
according to  RMT.  This is possibly due to the finite conductance in the
system\cite{FyoM94,Mir97}.  The dip at $r\sim 7$ with $t=30$ could be due to
insufficient averaging. Our current data do not permit to draw a definite
conclusion here, it could also represent a systematic effect.

Summarising our results for $r > l$, 
we have made the following observations.  First, in our
simulations, the spatial
structure of ALS depends on the dimension.
In  Q1D samples, we have observed
a global redistribution of the background intensity
towards the localisation centre 
(as compared with typical, uniformly spread
extended states).
In 3D samples, on the other hand, 
the background intensity is roughly constant; 
the wave-function intensity is significantly increased only in the
very narrow vicinity of the localisation centre.

Second, the DNLSM appears to describe
the spatial structure of background intensity of ALS adequately
in the Q1D Anderson model.
The agreement is good for all values of $t$ studied here. This verifies
the assumption that the semiclassical picture of a diffusive electron
accounts for the origin of ALS in the Q1D Anderson model. 

Third, we must emphasise that 
at least in the  Q1D case it appears to be  difficult to reach the asymptotic
regime where (\ref{eq:psi2asy}) is valid. We have thus not
been able to observe the  line shape suggested\cite{Mir97}
to be characteristic of ALS in Q1D samples. We expect, however, 
that this characteristic shape is {approached} further as $t$ is increased.

Fourth, fluctuations around the average ALS, in Q1D,  are consistent
with Eq.~(\ref{eq:psiq}). The agreement is the better the larger $t$ is, as
seen in Fig.~\ref{fig:varq1d}. This is so because Eq.~(\ref{eq:psiq})
was derived (in Ref.~\onlinecite{Mir97}) using        {asymptotic}
expressions. 

Fifth, in 3D samples the fluctuations around the average ALS appear
to be consistent with RMT, although the second moment is somewhat
larger than expected according to RMT.

Finally, we briefly consider the region $r < \ell$.
In the 3D case, the central peak is found to be very narrow. It is possibly 
much narrower than the electronic mean free path $\ell$  which is of
the order of several lattice spacings in the metallic
regime\cite{Nik02}.  This may indicate that the DNLSM is not
applicable in this case, since $\ell$ is the shortest relevant length
scale in the DNLSM.  It may also point towards the conjecture of
Refs.~\onlinecite{SmoA97} and \onlinecite{Mir00}  where  high
wave-function
amplitudes were suggested to arise from partial trapping of electrons
in rare local potential cavities. Such features are not included in
the DNLSM. At this point, however, we cannot offer quantitative
results  concerning the parametric dependence of the width of
the central peak.

In conclusion, 
we have studied the spatial 
structure and statistics of anomalously localised states in
the Anderson model.  Our results indicate~\cite{note} that the 
spatial structure of ALS
in Q1D and 3D samples is very different, as surmised in
Refs.~\onlinecite{Mir97,SmoA97}, and \onlinecite{Mir00}.  Our results are
consistent with the idea that the origin of ALS is different in Q1D and in
3D. In order to decide to which extent local potential traps are relevant in
3D, it would be of great interest to study the parametric dependence of the
width of the local maximum seen in Fig.~\ref{fig:als3d}.  In this context, a
continuous model would probably be more suitable than the discrete lattice
Hamiltonian (\ref{eq:defH}) studied in the present paper.

{\em Acknowledgement.} This work was supported by
a grant from Vetenskapsr\aa{}det, and by
the European Research Training Network QTRANS.


\begin{thebibliography}{10}
\bibitem{note}
 It cannot be excluded with certainty, however, that in the 3D case the
 most localised states are found only for some special
 configurations\cite{SmoA97}
 of the disorder potential which  may not have appeared
 in the - necessarily finite - ensemble of samples analysed here.
\end{thebibliography}
\end{document}